\documentclass[symmetry,article,accept,pdftex]{mdpi}
\usepackage{amsmath,latexsym}

\firstpage{1}
\pubvolume{xx}
\issuenum{1}
\articlenumber{5}
\pubyear{2020}
\copyrightyear{2020}
\history{Received: ; Accepted: ; Published: }

\Title{Lensing Effects in Retarded Gravity}
\Author{Asher Yahalom $^{1,2}$\orcidA{}}

\AuthorNames{Asher Yahalom}

\address{%
$^{1}$ \quad Ariel University, Faulty of Engineering, Department of Electrical \& Electronic Engineering, Ariel 40700, Israel; asya@ariel.ac.il\\

$^{2}$ \quad Ariel University, Center for Astrophysics, Geophysics, and Space Sciences (AGASS),
Ariel 40700, Israel;
}

\abstract{Galaxies are gigantic physical systems having a typical size of many tens of thousands of light years. Thus any change at the center of the galaxy will affect the rim only tens of millennia later. Those retardation effects seems to be ignored in present day modelling used to calculate rotational velocities of matter in the outskirts of the galaxy and the surrounding gas. The significant discrepancies between the velocities predicted by Newtonian theory and observed velocities are usually handled by either assuming an unobservable type of matter denoted "dark matter" or by modifying the laws of gravity (MOND as an example). Here we will show that considering general relativistic effects without neglecting retardation one can explain the apparent excess matter leading to gravitational lensing in both galaxies and galaxy clusters.}


\begin{document}

\newcommand{\beq} {\begin{equation}}
\newcommand{\enq} {\end{equation}}
\newcommand{\ber} {\begin {eqnarray}}
\newcommand{\enr} {\end {eqnarray}}
\newcommand{\eq} {equation}
\newcommand{\eqs} {equations }
\newcommand{\mn}  {{\mu \nu}}
\newcommand{\abp}  {{\alpha \beta}}
\newcommand{\ab}  {{\alpha \beta}}
\newcommand{\sn}  {{\sigma \nu}}
\newcommand{\rhm}  {{\rho \mu}}
\newcommand{\sr}  {{\sigma \rho}}
\newcommand{\bh}  {{\bar h}}
\newcommand{\br}  {{\bar r}}
\newcommand {\er}[1] {equation (\ref{#1}) }
\newcommand {\ern}[1] {equation (\ref{#1})}
\newcommand {\Ern}[1] {Equation (\ref{#1})}
\newcommand{\hdz}  {\frac{1}{2} \Delta z}

\section {Introduction}

Einstein's general relativity (GR) is known to be invariant under general coordinate modifications.
This group of general transformations has a Lorentz subgroup, which is valid even in the weak field approximation. This is seen through the field equations containing the d'Alembert (wave) operator, which can be solved using a retarded potential~solution.

It is known that GR is verified by many types of observations.
However, currently, Newton–Einstein gravitational theory is at a crossroads. It has much in its favor observationally, and it has some very disquieting challenges. The~successes that it has achieved in both astrophysical and cosmological scales have to be considered in light of the fact that GR needs to appeal to two unconfirmed ingredients, dark matter and energy, to achieve these successes. Dark matter has not only been with us since the 1920s (when it was initially known as the missing mass problem), but~it has also become severe as more and more of it had to be introduced on larger distance scales as new data have become available. Here we will be particularly concerned in the excess dark matter needed to justify observed gravitational lensing. Moreover, 40-year-underground and accelerator searches and experiments have failed to establish its existence. The~dark matter situation has become even more disturbing in recent years as the Large Hadron Collider was unable to find any super symmetric particle candidates, the community's preferred form of dark matter.

While things may still take turn in favor of the dark matter hypothesis, the~current situation is serious enough to consider the possibility that the popular paradigm might need to be amended in some way if not replaced altogether. The~present paper seeks such a modification. Unlike other ideas such as Milgrom's MOND
\cite{Mond}, Mannheim's Conformal Gravity \cite{Mannheim0,Mannheim1,Mannheim2},
Moffat's MOG \cite{MOG} or $f(R)$ theories and scalar-tensor gravity \cite{Corda}, the~present approach is, the~minimalist one adhering to the razor of Occam. It suggests to replace dark matter by effects within standard GR.

\begin{figure}[H]
\centering
\includegraphics[width=\columnwidth]{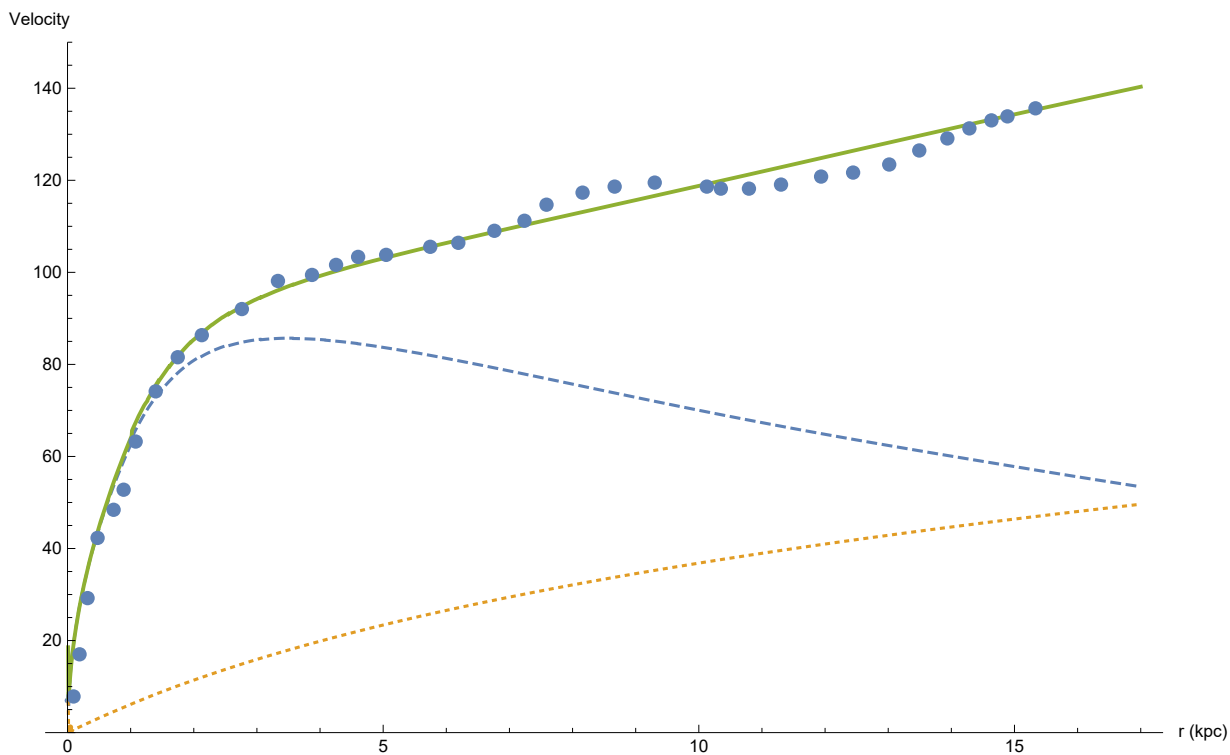}
 \caption{Velocity curve for M33. Observational points were obtained by Dr. Michal Wagman, a~former PhD student under~my supervision, using~\cite{Corbelli2}; the full line describes the  rotation curve, which is the sum of the dotted line, describing the retardation contribution, and~the dashed line, which is Newtonian.}
 \label{vcrhoc2}
\end{figure}

 Fritz Zwicky noticed in 1933 that the velocities of Galaxies within the Comma Cluster are much higher than those predicted by the virial calculation that assumes Newtonian theory~\cite{zwicky}.  He~calculated that the amount of matter required to
  account for the velocities could be 400 times greater with respect to that of visible matter, which led to suggesting  dark
   matter throughout the cluster. In~1959, Volders, pointed out  that  stars in the outer rims of the nearby galaxy M33
   do not move according to Newtonian theory~\cite{volders}. The virial theorem coupled with Newtonian Gravity implies that $MG/r \sim M v^2$, thus the~expected rotation curve should at some point decrease as $1/\sqrt{r}$.
   During the seventies Rubin and Ford~\cite{rubin1,rubin2} showed that, for~a large number of spiral galaxies, this behavior can be considered generic: velocities at the rim of the galaxies do not decrease— but they attain a plateau at some unique velocity, which differs for every galaxy. We have shown that such velocity curves can be deduced from GR if retardation is not neglected. The~derivation of the retardation force is described in previous publications~\cite{YaRe1,ge,YaRe2,Wagman,YaRe3,YaRe4,YaRe5}, see for example figure \ref{vcrhoc2}.

Previous work assumed a test particle moving slowly with respect to the speed of light as is appropriated for the case of galactic rotation curves, this is not adequate when the test particle is a photon moving in the speed of light as in the case relevant to gravitational lensing. Here, a different mathematical approach is needed as described in the current paper.

A gravitational lens is some form of matter (for example a cluster of galaxies) between a distant source of light and the observer, that is bending the light as it travels towards the observer.
\begin{figure}[H]
\centering
\includegraphics[width=0.7\columnwidth]{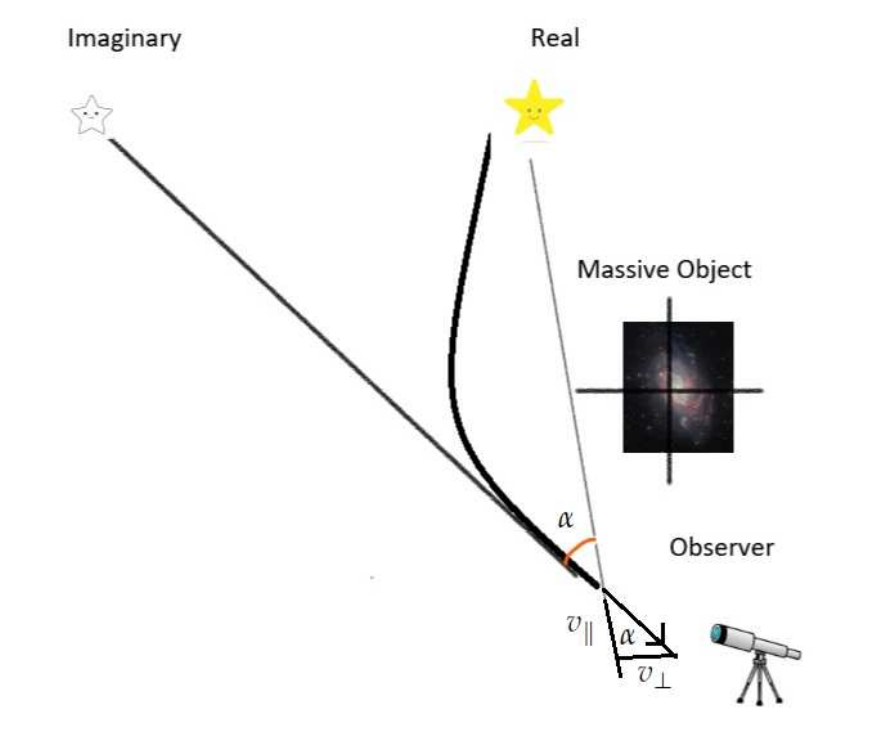}
 \caption{Light travelling toward the observer is bent due to the gravitational field of a massive object, thus a distant star appears to the observer at an angle $\alpha$ with resect to its true location. The tangent of alpha is the velocity of light perpendicular to the original direction
 $v_\bot$ divided by the velocity of light in the original direction $v_\|$ .}
 \label{lensifig}
\end{figure}

This effect is denoted gravitational lensing, the amount of bending is one of the predictions of GR \cite{[1],padma,[2]}.
It should be noted that Newtonian physics also predicts light bending, but only half of that predicted by GR \cite{[3]}.

Einstein made unpublished work on gravitational lensing as early as 1912 \cite{[4]}. In 1915 Einstein showed how GR explained the anomalous perihelion advance of the planet Mercury without any arbitrary parameters \cite{[11]}, in 1919 an expedition led by Eddington confirmed GR's prediction for the deflection of starlight by the Sun in the total solar eclipse of May 29, 1919,\cite{[12],[12a]} making Einstein famous \cite{[11]} instantly. The reader should recall that there was a special significance to a British scientist confirming the prediction of a German scientist after the bloody battles of world war I.

The fact that distortion of space time is proportional to the amount of mass that causes the distortion led to the use of gravitational lensing as a tool for proving the existence
of dark matter.

Strong lensing is the distortion of background galaxies into arcs when their light passes through a gravitational lens. This has been observed around many clusters such as Abell 1689 \cite{[53]}. By measuring the observed geometry, the mass of the in-between cluster can be calculated. In the cases where this was done, the mass-to-light ratios corresponded to the dynamical dark matter of clusters \cite{[54]}. Here we will show that is not a coincidence, and retardation dictates that this should be so.  Lensing can cause multiple copies of an image. By analyzing multiple image copies, astronomers have been able to map the distribution of dark matter around the MACS J0416.1-2403 galaxy cluster \cite{[55],[56]}.

Weak gravitational lensing is concerned with small distortions of galaxies, using statistical methods from huge galaxy surveys. By examining the shear deformation of the adjacent background galaxies, the distribution of dark matter can be calculated. The mass-to-light ratios correspond to dark matter densities predicted by other large-scale structure measurements \cite{[57]}.

We underline that dark matter does not bend light itself; mass (in this case the alleged mass of the dark matter) bends spacetime. Light follows the curvature of spacetime, resulting in the lensing effect \cite{[58],[59]}. Here we will show that the above described effects may be attributed to retardation and baryonic matter, with no additional form of (dark) matter.

The structure of the paper is as follows: First we describe the main results of general relativity, this is followed by the weak field approximation. Next we derive the geodesic equations for a particle moving at a speed of light ("a photon"). This is followed by a comparison of the current approach to that of Weinberg \cite{Weinberg}. The classical result of Einstein and Eddington are re-derived. Finally we discuss the effect of retardation on the lensing phenomena and show in what way can retardation cause a "dark matter" phenomena.

\section {General Relativity}

The general theory of relativity is based on two fundamental equations, the~first being Einstein equations~\cite{Narlikar,Weinberg,MTW,Edd}:
\beq
G_\mn = -\frac{8 \pi G}{c^4} T_\mn
\label{ein}
\enq
$G_\mn$ stands for the Einstein tensor (see \ern{Eint}), $T_\mn$ indicates the
stress–energy tensor  (see \ern{fltens}), $G \simeq 6.67 \ 10^{-11} \ \text{m}^3 \text{kg}^{-1} \text{s}^{-2}$ is the universal gravitational constant and $c \simeq 3 \ 10^8  \ \text{m s}^{-1}$ indicates the velocity of light in the absence of matter (Greek letters are indices in the range $0-3$).
The second fundamental equation that GR is based on is the geodesic equation:
\beq
\frac{d^2 x^\alpha}{dp^2}+\Gamma^\alpha_\mn \frac{d x^\mu}{dp} \frac{d x^\nu}{dp} = \frac{d u^\alpha}{dp}+\Gamma^\alpha_\mn u^\mu u^\nu = 0
\label{geo}
\enq
$x^\alpha (p)$ are the coordinates of the particle in spacetime, $p$ is a typical parameter along the trajectory that for massive particles is chosen to be the length of the trajectory ($p=s$), $u^\mu = \frac{d x^\mu }{d p}$  is the $\mu$-th component of the 4-velocity of a massive particle moving along the geodesic trajectory  (increment of $x$ per $p$) and $\Gamma^\alpha_\mn$ is the affine connection (Einstein summation convention is assumed). The~stress–energy tensor of matter is usually taken in the form:
\beq
T_\mn = (pr+\rho c^2) u_\mu  u_\nu - pr \ g_\mn
\label{fltens}
\enq

In the above, $pr$ is the pressure and $\rho$ is the {\bf mass} density. We remind the reader that lowering and raising
indices is done through the metric $g_\mn$ and  inverse metric $g^\mn$, such that $u_\mu= g_\mn u^\nu$. The~same metric serves to calculate $s$:
\beq
ds^2 = g_\mn dx^\mu  dx^\nu,
\label{intervale}
\enq
for a photon $ds=0$. The affine connection is connected to the metric as follows:
\beq
\Gamma^\alpha_\mn = \frac{1}{2} g^\ab \left(g_{\beta \mu, \nu} + g_{\beta \nu, \mu} - g_{\mn, \beta}\right), \qquad
 g_{\beta \mu, \nu} \equiv \frac{\partial g_{\beta \mu}}{\partial x^\nu}
\label{affine}
\enq
Using the affine connection we calculate the Riemann and Ricci tensors and the curvature scalar:
\beq
R^\mu_{\nu \ab} = \Gamma^\mu_{\nu \alpha,\beta} - \Gamma^\mu_{\nu \beta,\alpha} + \Gamma^\sigma_{\nu \alpha} \Gamma^\mu_{\sigma \beta}
- \Gamma^\sigma_{\nu \beta} \Gamma^\mu_{\sigma \alpha}, \quad R_{\ab}= R^\mu_{\ab \mu}, \quad R= g^\ab R_{\ab}
\label{RieRicci}
\enq
which, in~turn, serves to calculate the Einstein tensor:
\beq
G_{\ab}= R_{\ab} - \frac{1}{2} g_\ab R.
\label{Eint}
\enq
Hence, the~given matter distribution determines the metric through Equation~(\ref{ein}) and the metric determines the geodesic trajectories through Equation~(\ref{geo}). Those equations are well known to be symmetric under smooth coordinate transformations (diffeomorphism).
\beq
x'_{\alpha}= x'_{\alpha} (x_\mu).
\label{Gct}
\enq

\section {Linear Approximation of GR - The Metric}
\label{linear}

Only in extreme cases of compact objects (black holes and neutron stars) and the primordial reality or the very early universe does one need not consider the solution of the full non-linear Einstein Equation~\cite{YaRe1}. In~typical cases of astronomical interest (the galactic case included) one can use a linear approximation to those equations around the flat Lorentz metric $\eta_{\mn}$ such that (Private communication with the late Professor Donald Lynden-Bell):
 \beq
 g_{\mn} = \eta_{\mn} + h_{\mn}, \quad \eta_{\mn} \equiv \ {\rm diag } \ (1,-1,-1,-1),
 \qquad
 |h_{\mn}|\ll 1
 \label{lg}
 \enq
 One then defines the quantity:
 \beq
 \bar h_\mn \equiv h_\mn -  \frac{1}{2} \eta_\mn h, \quad h = \eta^{\mn} h_{\mn},
 \label{bh}
 \enq
 $\bar h_\mn = h_\mn $ for non diagonal terms. For~diagonal terms:
 \beq
 \bar h = - h \Rightarrow  h_\mn = \bar h_\mn -  \frac{1}{2} \eta_\mn \bar h .
 \label{bh2}
 \enq

  The general coordinate transformation symmetry of Equation~(\ref{Gct}) has a subgroup  of infinitesimal transformations which are manifested in the gauge freedom of $h_\mn$ in the weak field approximation. It~can be shown (\cite{Narlikar} page 75, exercise 37, see also~\cite{Edd,Weinberg,MTW}) that one can  choose a gauge such that the Einstein equations are:
 \beq
\Box \bh_{\mn} \equiv \bh_{\mn, \alpha}{}^{\alpha}=-\frac{16 \pi G}{c^4} T_\mn , \qquad \bh_{\mu \alpha,}{}^{\alpha}=0.
\label{lineq1}
\enq

The d'Alembert operator $\Box$ is clearly invariant under the Lorentz symmetry group (another subgroup of the general coordinate transformation symmetry described by Equation (\ref{Gct})), of~which the Newtonian Laplace operator
$\vec \nabla^2$ is not, but~this comes with the price that "action at a distance" solutions are forbidden and only retarded solutions are
allowed. The~$T_\mn$ stress energy tensor should be calculated at the appropriate frame and thus for the massive body causing the lensing, matter is approximately at~rest.

\Ern{lineq1} can always be integrated to take the form~\cite{Jackson}:
 \ber
& & \bh_{\mn}(\vec x, t) = -\frac{4 G}{c^4} \int \frac{T_\mn (\vec x', t-\frac{R}{c})}{R} d^3 x',
\nonumber \\
 t &\equiv& \frac{x^0}{c}, \quad \vec x \equiv x^a \quad a,b \in [1,2,3], \nonumber \\
  \vec R &\equiv& \vec x - \vec x', \quad R= |\vec R |.
\label{bhint}
\enr
For reasons why the symmetry between space and time is broken, see~\cite{Yahalom,Yahalomb}.
The factor before the integral is small: $\frac{4 G}{c^4} \simeq 3.3 \times 10^{-44}$ in MKS units; hence, in~the above calculation one can take $T_\mn$, which is zero order in $h_\abp$.

In~the zeroth order:
\beq
u^0=\frac{1}{\sqrt{1-\frac{v^2}{c^2}}}, \quad u^a = \vec u =\frac{\frac{\vec v}{c}}{\sqrt{1-\frac{v^2}{c^2}}} , \quad v^\mu \equiv  \frac{d x^\mu}{d t}, \quad
\vec v \equiv  \frac{d \vec x}{d t}, \quad v= |\vec v|,
\label{uz}
\enq
in which we assume that massive body causing the lensing effect is composed of massive particles
(those equations will not be correct for the photon which is affected by the lensing potential).

Assuming the reasonable assumption that the said massive body is composed of particles of non relativistic velocities:
\beq
u^0 \simeq 1,  \qquad \vec u \simeq \frac{\vec v}{c} , \qquad u^a \ll u^0   \qquad {\rm for} \quad v \ll c.
\label{uzslo}
\enq
Let us now look at equation~(\ref{fltens}). We~assume $\rho c^2 \gg pr$ and, taking into account equation~(\ref{uzslo}), we~arrive at $T_{00} = \rho c^2 $, while  other tensor components are significantly smaller. Thus, $\bar h_{00}$ is significantly larger than other components of $\bar h_\mn$ which are ignored from now on. One should notice that it is  possible to deduce from the  gauge condition in equation~(\ref{lineq1}) the relative order of magnitude of the relative components of $h_\mn$:
\beq
\bar h_{\alpha 0,}{}^0=-\bar h_{\alpha a,}{}^a \qquad \Rightarrow
\bar h_{00,}{}^0=-\bar h_{0 a,}{}^a, \quad \bar h_{b0,}{}^0=-\bar h_{b a,}{}^a.
\label{gaugeim}
\enq

Thus, the~zeroth derivative of $\bar h_{00}$ (which contains a $\frac{1}{c}$ as $x^0 = c t$) is the same order as the spatial derivative
of $\bar h_{0a}$  meaning that $\bar h_{0a}$ is of order $\frac{v}{c}$ smaller than $\bar h_{00}$.
And the zeroth derivative of $\bar h_{0a}$ (which appears in Equation~(\ref{gaugeim})) is the same order as the spatial derivative of $\bar h_{ab}$. Meaning that $\bar h_{ab}$ is of order $\frac{v}{c}$ with respect to $\bar h_{0a}$ and of order $(\frac{v}{c})^2$ with respect to $\bar h_{00}$.

In the current approximation, the following results hold:
 \beq
 \bar h = \eta^\mn \bar h_\mn = \bar h_{00}.
 \label{bh5}
 \enq
 \beq
 h_{00} = \bar h_{00} -  \frac{1}{2} \eta_{00} \bar h = \frac{1}{2} \bar h_{00} .
 \label{bh6}
 \enq
 \beq
 h_{\underline{aa}} = -  \frac{1}{2} \eta_{\underline{aa}} \bar h = \frac{1}{2} \bar h_{00} .
 \label{bh7}
 \enq
 (The underline $\underline{aa}$ signifies that the Einstein summation convention is not assumed).
\beq
 h_{\mn} = \bar h_{\mn} = 0, \qquad \mu \neq \nu .
 \label{bh8}
 \enq
We can summarize the above results in a concise formula:
\beq
 h_{\mn} = h_{00} \delta_{\mn}.
 \label{bh9}
  \enq
in which $\delta_{\mn}$ is Kronecker's delta. It will be useful to introduce the gravitational potential $\phi$  which is defined below and can be calculated using Equation~(\ref{bhint}):
\beq
\phi \equiv \frac{c^2}{4} \bar h_{00}
= -\frac{ G}{c^2} \int \frac{T_{00} (\vec x', t-\frac{R}{c})}{R} d^3 x'
= -G \int \frac{\rho (\vec x', t-\frac{R}{c})}{R} d^3 x'
\label{phi}
\enq
from the above definition and \ern{bh6} it follows that:
\beq
 h_{00} = \frac{2}{c^2} \phi.
 \label{phih00}
  \enq

\section {Linear Approximation of GR - The Lensing Trajectory}

Let us start calculating the lensing trajectory by writing $u^\mu = \frac{d x^\mu }{d p}$ in terms of the notation introduced in \ern{bhint}:
\beq
u^0 = \frac{d x^0 }{d p} = c   \frac{d t }{d p}, \qquad
u^a = \frac{d x^a }{d p} = \frac{d x^a }{d t}   \frac{d t }{d p} = v^a \frac{d t }{d p}
\label{uint}
\enq
it thus follows that:
\beq
\frac{d u^a }{d p} = \frac{d v^a }{d p} \frac{d t }{d p}+ v^a \frac{d^2 t }{d p^2}
\label{uints}
\enq
Taking into account \ern{geo} we obtain:
\beq
 \frac{d v^a }{d p} \frac{d t }{d p}+ v^a \frac{d^2 t }{d p^2} = - \Gamma^\alpha_\mn u^\mu u^\nu
\label{uints2}
\enq
multiplying by $\left(\frac{d p}{d t}\right)^2$ and using the notation of \ern{uz} we obtain:
\beq
 \frac{d v^a }{d t} + v^a \left(\frac{d p}{d t}\right)^2 \frac{d^2 t }{d p^2} = - \Gamma^\alpha_\mn v^\mu v^\nu.
\label{uints3}
\enq
However, according to \ern{geo}:
\beq
  \frac{d^2 t }{d p^2} = \frac{1}{c} \frac{d^2 x^0 }{d p^2}  =  -  \frac{1}{c}  \Gamma^0_\mn u^\mu u^\nu.
\label{uints4}
\enq
Inserting \ern{uints4} into \ern{uints3} we arrive at the form:
\beq
 \frac{d v^a }{d t}  = - \Gamma^\alpha_\mn v^\mu v^\nu + \Gamma^0_\mn v^\mu v^\nu  \frac{v^a}{c}.
\label{uints5}
\enq
Let us now calculate the affine connection in the linear approximation:
\beq
\Gamma^\alpha_\mn = \frac{1}{2} \eta^\abp \left(h_{\beta \mu, \nu} + h_{\beta \nu, \mu} - h_{\mn, \beta}\right).
\label{affinel}
\enq
Taking into account \ern{affinel} and \ern{bh9} we obtain:
\beq
\Gamma^a_\mn v^\mu v^\nu =
 \eta^{a \beta} \left(h_{\beta \mu, \nu} v^\mu v^\nu  -  \frac{1}{2}h_{\mn, \beta}v^\mu v^\nu \right)
 = \eta^{a \beta} \left(\delta_{\beta \mu} \partial_\nu h_{00} v^\mu v^\nu  - \frac{1}{2} \delta_{\mn} \partial_\beta h_{00} v^\mu v^\nu \right).
\label{affinel2}
\enq
The affine connection has only first order terms in $h_\abp$; hence, to~the first order
$\Gamma^\alpha_\mn v^\mu v^\nu$ appearing in the geodesic,  $v^\mu v^\nu$ is of the zeroth order.
By definition $v^0 = \frac{d x^0}{dt} = c$. Also the null interval of the photon is given to zeroth order in $h_{00}$ (that is in the absence of a gravitational field) by:
\beq
0=ds^2 = \eta_{\mn} dx^\mu dx^\nu \Rightarrow
0= \eta_{\mn} v^\mu v^\nu  = (v^0)^2 - (\vec v)^2 \Rightarrow v = c.
\label{nulgeo}
\enq
It follows that in the zeroth order approximation:
\beq
\delta_{\mn} v^\mu v^\nu = (v^0)^2 + \vec{v}^2 = 2 c^2
\label{affinel3}
\enq
and also in the same approximation:
\beq
\eta^{a \beta} \delta_{\beta \mu} v^\mu v^\nu = \eta^{a \mu} v^\mu v^\nu = - v^a v^\nu.
\label{affinel4}
\enq
We now combine the above results and write:
\beq
\Gamma^a_\mn v^\mu v^\nu = - v^a v^\nu \partial_\nu h_{00}  - c^2 \partial^a h_{00}.
\label{affinel5}
\enq
We now turn our attention to the second term in \ern{uints5}
\beq
\Gamma^0_\mn v^\mu v^\nu  \frac{v^a}{c} =
\eta^{0 \beta} \left(h_{\beta \mu, \nu} v^\mu v^\nu  -  \frac{1}{2}h_{\mn, \beta}v^\mu v^\nu \right)
\frac{v^a}{c} =
\left(h_{0 \mu, \nu} v^\mu v^\nu  -  \frac{1}{2}h_{\mn, 0}v^\mu v^\nu \right) \frac{v^a}{c}
\label{secgeo1}
\enq
inserting \ern{bh9} into \ern{secgeo1} we arrive at the result:
\beq
\Gamma^0_\mn v^\mu v^\nu  \frac{v^a}{c} =
\left(\delta_{0 \mu} \partial_\nu h_{00} v^\mu v^\nu  -  \delta_{\mn}
\frac{1}{2} \partial_0 h_{00} v^\mu v^\nu \right) \frac{v^a}{c} =
\left(c v^\nu \partial_\nu h_{00} -  c^2 \partial_0 h_{00} \right) \frac{v^a}{c} =
v^a v^b \partial_b h_{00}.
\label{secgeo2}
\enq
Combining the results from \ern{affinel5} and \ern{secgeo2} into \ern{uints5} we arrive at the
photon's equation of motion:
\beq
 \frac{d v^a }{d t}  = c^2 \partial^a h_{00} + v^a \left[v^\nu \partial_\nu h_{00} + v^b \partial_b h_{00} \right].
\label{photonequa}
\enq
In terms of the gravitational potential  this can be rewritten using \ern{phih00} as follows:
\beq
 \frac{d \vec v }{d t}  = -c^2 \vec \nabla h_{00} + 2 \vec v \left(\vec v \cdot \vec \nabla h_{00} \right) + \vec v \partial_t h_{00} \quad
 \Rightarrow \quad
 \frac{d \vec v }{d t}  = -2 \vec \nabla \phi + 4 \frac{\vec v}{c} \left(\frac{\vec v}{c} \cdot \vec \nabla \phi \right) + 2 \frac{\vec v}{c^2} \partial_t \phi.
\label{photonequa2}
\enq
This equation is almost identical to equation 9.2.6 of Weinberg \cite{Weinberg}. Notice, however, that Weinberg neglects the term $2 \frac{\vec v}{c^2} \partial_t \phi$ in his post Newtonian approximation. Although this term can be neglected or shown to be small in specific circumstances
such as case that $\phi$ is static or slowly varying, its removal may lead to inconsistencies as we explain below.

Let us inspect the null interval \ern{intervale}:
\beq
ds^2 = g_\mn dx^\mu  dx^\nu = 0,
\label{intervale2}
\enq
taking into account \ern{lg} and \ern{bh9} this takes the form:
\beq
 0 = ds^2 = (\eta_{\mn} + h_{\mn}) dx^\mu  dx^\nu = (\eta_{\mn} +  h_{00} \delta_{\mn}) dx^\mu  dx^\nu,
\label{intervale3}
\enq
Hence:
\beq
 0 =  (\eta_{\mn} +  h_{00} \delta_{\mn}) v^\mu  v^\nu \Rightarrow
 (1+ h_{00}) c^2 - (1-h_{00}) \vec v^2 = 0 \Rightarrow  v^2 = c^2 \frac{(1+ h_{00})}{(1-h_{00})}
\label{intervale4}
\enq
Or:
\beq
v = c \sqrt{\frac{(1+ h_{00})}{(1-h_{00})}} = c \left((1+ h_{00} + O (h_{00}^2)\right)
\label{intervale5}
\enq
compare to Weinberg \cite{Weinberg} equation 9.2.5. Since by \ern{phi} and \ern{phih00} $h_{00}$
is a small negative number, it follows that $v<c$ (notice that this result holds for a global coordinate system, in the local flat coordinate system the velocity will of course be exactly $c$).
Now:
\beq
  v \frac{dv}{dt} = \vec v \cdot \frac{d \vec v}{dt}
\label{consis1}
\enq
To first order in $h_{00}$:
\beq
  v \frac{dv}{dt} = c^2  \frac{d h_{00}}{dt}.
\label{consis2}
\enq
Taking into account \ern{photonequa2}:
\beq
 \vec v  \cdot \frac{d \vec v}{dt} = \vec v  \cdot \left[ -c^2 \vec \nabla h_{00} + 2 \vec v \left(\vec v \cdot \vec \nabla h_{00} \right) + \vec v \partial_t h_{00}\right]
 = -c^2 \vec v  \cdot \vec \nabla h_{00}+ 2 v^2 \left(\vec v \cdot \vec \nabla h_{00} \right) +  v^2 \partial_t h_{00}
\label{consis3}
\enq
Thus to the same first order of $h_{00}$ we have:
\beq
 \vec v  \cdot \frac{d \vec v}{dt}
 =  c^2 \left(\vec v \cdot \vec \nabla h_{00} \right) +  c^2 \partial_t h_{00}
 = c^2  \frac{d h_{00}}{dt}
\label{consis4}
\enq
which is dependent on not neglecting the $\partial_t h_{00}$ term. Thus this term
is absolutely necessary to maintain the identity of \ern{consis1} and without it we
are led to a contradiction. We conclude that for a time dependent gravitational potential the
$\partial_t h_{00}$ term is required.

Let us consider a photon travelling at a straight line in the direction $\hat v_0$ in the absence of gravity, the velocity of this photon would be $\vec v_0 = c \hat v_0$. Now let us assume that the photon passes near a weak gravitational source such that \ern{photonequa2} is valid. We thus decompose the photon velocity field to two components one which parallel and one which is perpendicular to its original direction:
\beq
 \vec v_\| =  (\vec v \cdot \hat v_0) \hat v_0, \quad \vec v_\bot =  \vec v -  \vec v_\|
 = \vec v - (\vec v \cdot \hat v_0) \hat v_0, \quad \vec v =\vec v_\bot +  \vec v_\|
\label{vpar}
\enq
Thus it follows from \ern{photonequa2} that:
\beq
 \frac{d v_\| }{d t}  =
 \frac{d (\hat v_0 \cdot \vec v)}{d t}  =
    -c^2 \hat v_0 \cdot \vec \nabla h_{00} + 2 \hat v_0 \cdot \vec v \left(\vec v \cdot \vec \nabla h_{00} \right) + \hat v_0 \cdot \vec v \partial_t h_{00}
 \label{photonequapar1}
\enq
Thus to first order in $h_{00}$:
\beq
 \frac{d v_\| }{d t}  =
     -c \vec v \cdot \vec \nabla h_{00} + 2 c \left(\vec v \cdot \vec \nabla h_{00} \right) + c \partial_t h_{00} = c \left( \partial_t h_{00} + \vec v \cdot \vec \nabla h_{00} \right)=
     c \frac{d  h_{00} }{d t}
 \label{photonequapar2}
\enq
in which we remember that $\vec v = \vec v_0 + O(h_{00}) = c \hat v_0 + O(h_{00})$.
This can be integrated as follows:
\beq
  v_\| = c h_{00} + k
 \label{photonequapar3}
\enq
in which $k$ is a constant. Far away from the gravitational source
$\lim_{|\vec x|->\infty} v_\| = c, \quad \lim_{|\vec x|->\infty} h_{00} = 0 $,
hence $k=c$, and we may write:
\beq
  v_\| = c (h_{00} + 1).
 \label{photonequapar4}
\enq
Now:
\beq
  v_\|^2 = c^2 (1 + 2 h_{00}) + O (h_{00}^2).
 \label{photonequapar5}
\enq
And according to \ern{intervale5}:
\beq
  v^2 = c^2 (1 + 2 h_{00}) + O (h_{00}^2).
 \label{photonequaparb5}
\enq
It now follows from \ern{vpar}:
\beq
v_\bot^2 = v^2 -  v_\|^2 =   O (h_{00}^2) \quad \Rightarrow  \quad v_\bot = O (h_{00}).
\label{vpar2}
\enq
The lensing angle is defined (see figure \ref{lensifig})
\beq
\tan \alpha = \frac{v_\bot}{v_\|} = \frac{v_\bot}{c} +O(h_{00}^2)
\Rightarrow  \alpha \simeq \frac{v_\bot}{c}
\label{lensangle}
\enq
as the angle $\alpha$ is small, thus for lensing in the linear approximation only the perpendicular component $v_\bot$ is important. Now to the first order in $h_{00}$ we may write \ern{photonequa2} as:
\beq
 \frac{d \vec v }{d t}  =
  -c^2 \left(\vec \nabla h_{00} - \hat v_0 (\hat v_0 \cdot \vec \nabla h_{00})\right)
  +  \vec v_0 \left(\vec v \cdot \vec \nabla h_{00}  + \partial_t h_{00} \right)=
  -c^2 \vec \nabla_\bot h_{00} +  \vec v_0 \frac{d  h_{00} }{d t}
 \label{photonequa2c}
\enq
in which we define the perpendicular gradient as: $\vec \nabla_\bot \equiv \vec \nabla  - \hat v_0 (\hat v_0 \cdot \vec \nabla)$.
Now taking into account \ern{vpar} and \ern{photonequapar2} we arrive at the result:
\beq
 \frac{d \vec v_\bot }{d t}  = \frac{d (\vec v - \vec v_\|) }{d t}=
  -c^2 \vec \nabla_\bot h_{00} =  -2 \vec \nabla_\bot \phi
 \label{photonequa2d}
\enq
in which we have used \ern{phih00} to give the perpendicular acceleration in terms of a gravitational potential.

\section{Other approaches to the problem of lensing}

Another approach to the problem of lensing than the one given above is to start from a  Schwarzschild metric as is done by Weinberg \cite{Weinberg}. This metric describe a static spherically symmetric mass distribution and thus is less general than the approach taken in this paper. It does have one advantage, however, and this is the ability to take into account strong gravitational fields and not
just weak ones. This advantage is irrelevant in most astronomical cases in which gravity is weak and must be only considered for light trajectories near compact objects (black holes \& neutron stars).
The Schwarzschild squared interval can be written as:
\beq
ds_{Schwarzschild}^2=\left(1-{\frac {r_s}{r'}}\right)c^{2} dt^{2}-\left(1-{\frac {r_s}{r'}}\right)^{-1}dr'^{2}
-r'^{2}\left(d\theta ^{2}+\sin ^{2}\theta \,d\phi ^{2}\right)
\label{Schwa}
\enq
In which $r',\theta,\phi$ are  spherical coordinates and the point massive body
is located at $r'=0$. The Schwarzschild radius is defined as:
\beq
r_s = \frac{2 G M}{c^2}
\label{Schwaradius}
\enq
in which $M$ is the mass of the point particle. Comparing the $g_{00}$ component of \ern{Schwa}
and \ern{intervale3} it follows that we can identify:
\beq
h_{00}= - \frac{r_s}{r'}
\label{hSchwa}
\enq
provided $\frac{r_s}{r'} \ll 1$ in accordance with \ern{lg}, hence the two approaches should coincide for:
\beq
r' \gg r_s
\label{hSchwa2}
\enq
Going back to \ern{intervale3} we have:
\beq
ds^2 = (\eta_{\mn} + h_{\mn}) dx^\mu  dx^\nu = (\eta_{\mn} +  h_{00} \delta_{\mn}) dx^\mu  dx^\nu = (1 +  h_{00}) c^2 dt^2 - (1 -  h_{00}) d\vec x^2 ,
\label{intervale3b}
\enq
it tempting to write $d\vec x^2 = dr^2 + r^2 \left(d\theta ^{2}+\sin^2 \theta d\phi ^{2}\right)$ as is usually done for spherical coordinates. But notice that $r$ is not strictly a radial coordinate which is defined as the circumference, divided by $2 \pi$, of a sphere centered around the massive body. In fact from \ern{intervale3b} it is clear that the appropriate radial coordinate is:
\beq
r' = r \sqrt{1-  h_{00}}
\label{rp}
\enq
which is a small correction to $r$. Now calculating the differential $dr'$ it follows that:
\beq
dr' = dr \sqrt{1-  h_{00}} + r d \sqrt{1-  h_{00}} = dr \frac{1-h_{00} - \frac{1}{2} r \frac{d h_{00}}{dr}}{\sqrt{1- h_{00}}}.
\label{drp}
\enq
According to \ern{hSchwa}:
\beq
h_{00}= - \frac{r_s}{r'} =- \frac{r_s}{r \sqrt{1-  h_{00}} }
\Rightarrow  h_{00} \sqrt{1-  h_{00}} = - \frac{r_s}{r} \Rightarrow  h_{00} = - \frac{r_s}{r}
\label{hSchwa3}
\enq
where the last equality is correct to first order. Alternatively one can use \ern{phi} to calculate the gravitational potential for a static point mass to obtain:
\beq
\phi = -G \frac{M}{r}
\label{phipoin}
\enq
and then plug this into \ern{phih00} to obtain again:
\beq
 h_{00} = \frac{2}{c^2} \phi = -\frac{2 G M}{c^2 r} = -\frac{r_s}{r}.
 \label{phih00poi}
  \enq
It now follows that:
\beq
\frac{d h_{00}}{dr} = \frac{r_s}{r^2} = - \frac{h_{00}}{r}
\label{dh00poi}
  \enq
Plugging \ern{dh00poi} into \ern{drp} leads to:
\beq
dr' = dr \frac{1-h_{00} - \frac{1}{2} r \frac{d h_{00}}{dr}}{\sqrt{1- h_{00}}}
= dr \frac{1- \frac{1}{2}h_{00}}{\sqrt{1- h_{00}}}.
\label{drp2}
\enq
Hence to first order in $h_{00}$:
\beq
dr' = dr.
\label{drp3}
\enq
Using the results \ern{rp} and \ern{drp3} the interval given in \ern{intervale3b} can be rewritten as:
\beq
ds^2 = (1 +  h_{00}) c^2 dt^2 - (1 -  h_{00}) d\vec x^2 =
(1 +  h_{00}) c^2 dt^2 - (1 -  h_{00}) dr'^2 - r'^2 \left(d\theta ^{2}+\sin^2 \theta d\phi ^{2}\right).
\label{intervale4b}
\enq
As to first order in $h_{00}$:
\beq
 1 -  h_{00} = \frac{1}{1 +  h_{00}},
\label{intervale5b}
\enq
and taking into account \ern{hSchwa} we obtain:
\beq
ds^2 = (1 - \frac{r_s}{r'}) c^2 dt^2 - (1 -\frac{r_s}{r'})^{-1}  dr'^2 - r'^2 \left(d\theta ^{2}+\sin^2 \theta d\phi ^{2}\right) = ds_{Schwarzschild}^2 .
\label{intervale6b}
\enq
Thus to first order our metric is identical to Schwarzschild's for the case of a static point particle. This makes our analysis superior as it addresses the case of a general density distribution and does not ignore the possibility of time dependence which is crucial for
retardation effects to take place.

 \section{Lensing in the static case}

Newtonian theory dictates that a body with any mass (since inertial mass and gravitational mass are equal) moving under the influence of gravity alone must follow a trajectory which is dictated by the
equation:
\beq
 \frac{d \vec v}{dt} = - \vec \nabla \phi_N.
\label{geol2}
\enq
The Newtonian potential $\phi_N$ resembles $\phi$ given by \ern{phi} but neglects the retardation effect such that:
\beq
\phi_N = -G \int \frac{\rho (\vec x', t)}{R} d^3 x'
\label{phiN}
\enq
In \cite{YaRe3} we have shown that the geodesic equations reduce for slow moving test particles to:
\beq
 \frac{d \vec v}{dt} = - \vec \nabla \phi.
\label{geol2b}
\enq
this will coincide with \ern{geol2} for a static density distribution or for a slowly changing mass distribution as is well known. For light rays we obtained  \ern{photonequa2d}, this also resembles \ern{geol2} but carries some major differences even for a static mass distribution. First there is a factor $2$ multiplying the potential which is missing in \ern{geol2}. Second this equation only describe motion perpendicular to the original direction of the light ray and not the propagation in the direction of the light ray itself.

Suppose the gravitational field is orthogonal to the light ray at least when the light ray is moving in proximity to the gravitating body, that is when the gravitational force is most significant.
In this case we can write approximately:
\beq
 \frac{d \vec v_\bot }{d t}  \simeq   -2 \vec \nabla \phi.
 \label{photonequa2d2}
 \enq
 It is tempting to make a further step and write:
 \beq
 \frac{d \vec v }{d t}  \simeq -2 \vec \nabla \phi
 \label{photonequa2d3}
 \enq
but the parallel component of $\vec v$ satisfies \ern{photonequapar4} and thus \ern{photonequa2d3}
is not correct even to the first order in $h_{00}$ at it simply states that this component is fixed
(remember we assume the force to be orthogonal to the light ray).
Notice, however, that according to \ern{lensangle} the lensing phenomena is not affected by the
parallel component, hence, no harm is done if we calculate this component to zero order instead of
first order. It follows that for a static mass distribution it suffices for the purpose of deducing the lensing effect to solve the equation:
 \beq
 \frac{d \vec v }{d t} \simeq -2 \vec \nabla \phi_N
 \label{photonequa2d4}
 \enq
which is just the Newtonian trajectory but with double the gravitational force. Now, the trajectory
of gravitating test particles in a Newtonian point mass gravitational field is known and is given
by the equation (Goldstein \cite{Goldst} equation 3.55) :
 \beq
 \frac{1}{r} = \frac{k}{l^2}[1+e \cos (\theta - \theta')]
 \label{Goldst}
 \enq
in which $r,\theta$ are standard cylindrical coordinates. A static point gravitating body is assumed located at the origin of axis and generating a gravitational potential of the form:
\beq
\phi =  -\frac{k}{r}.
\label{phiGoldst}
 \enq
 The quantity $k$ according to Newtonian theory is $k_N = G M$ which follows from
 \ern{geol2} and \ern{phipoin}. This would also be the value according to General Relativity
 for a slowly moving test particle (see \ern{geol2b}). However, for a light ray we should use
 \ern{photonequa2d4} instead and thus $k_E = 2 k_N = 2 G M$.

\begin{figure}[H]
\centering
\includegraphics[width=\columnwidth]{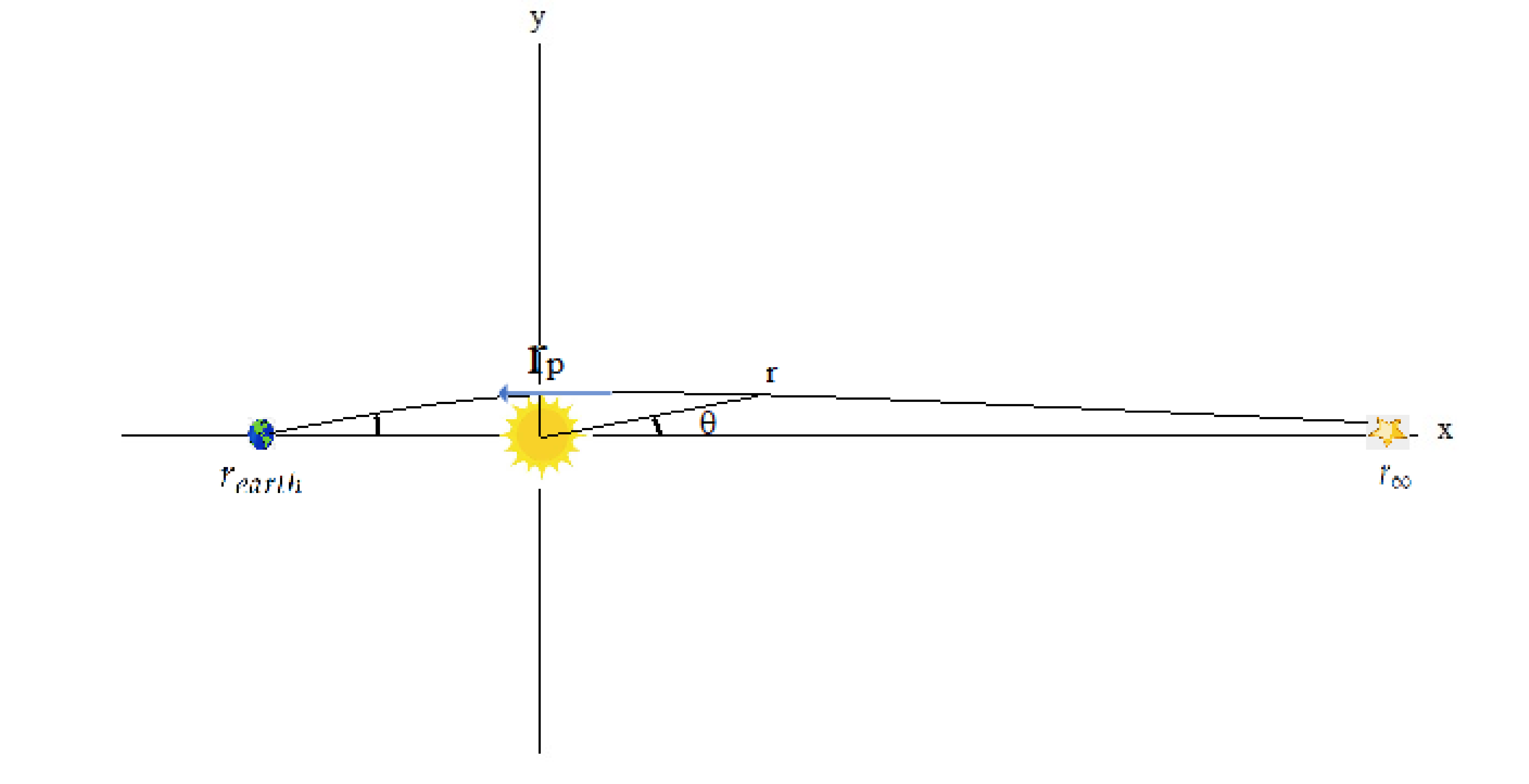}
 \caption{The trajectory of light from a distant star to the earth as affected by the sun's gravity}
 \label{tradiststar}
\end{figure}

 The quantity $l$ is a constant of motion (which can be thought as an angular momentum in the direction perpendicular to the plane of motion per unit mass) and is given by:
\beq
l = r^2 \dot{\theta}.
\label{angmom}
 \enq
The eccentricity $e$ is given by the equation:
\beq
e = \sqrt{1+\frac{2 E l^2}{k^2}}.
\label{ecc}
 \enq
 which is dependent on another constant of motion $E$ (which can be thought as the energy of the test particle per unit mass):
 \beq
E = \frac{1}{2} v^2 + \phi.
\label{energy}
 \enq
A light ray coming from far away will have $v=c$ at infinity while $\lim_{r->\infty} \phi = 0$ hence
we may write:
 \beq
E = \frac{1}{2} c^2.
\label{energyc}
 \enq
Thus:
\beq
e = \sqrt{1+\frac{c^2 l^2}{k^2}}.
\label{ecc2}
 \enq
Consider a star is located far away from the sun at a distance $r_\infty$. Let the line connecting this star and the sun which is located at $r=0$ coincide with the x-axis. Assume that the earth is also located along the x-axis and has an x coordinate of $x_{earth} = -r_{earth}$ as depicted in
figure \ref{tradiststar}.
We consider a light ray orbit in which a star emits a photon which is detected on earth. At the
closest point of the trajectory to the sun the photons travels at a direction which is purely in
the $\hat \theta$ direction and has no radial component. At this point the photon is a distance $r_p$ from the origin of axis and thus:
\beq
\vec v = r_p \dot{\theta_p} \hat \theta \Rightarrow  r_p \dot{\theta_p} = v \simeq c.
\label{vp}
 \enq
This is inserted into \ern{angmom} to yield:
\beq
l = r_p^2 \dot{\theta_p}  \simeq c r_p .
\label{angmom2}
 \enq
Hence:
\beq
e = \sqrt{1+\frac{c^4 r_p^2}{k^2}}.
\label{ecc222}
 \enq
Now according to general relativity $k=k_E = 2 G M$:
\beq
e_E = \sqrt{1+\frac{c^4 r_p^2}{4 G^2 M^2}}= \sqrt{1+\frac{r_p^2}{r_s^2}} \simeq \frac{r_p}{r_s}
\gg 1.
\label{eccE2}
 \enq
In which $r_s$ is the Schwarzschild radius defined in \ern{Schwaradius} and we assume the
weak field approximation as described in \ern{hSchwa2}, for the Newtonian theory we
have $k=k_N = G M$ for which we obtain:
\beq
e_N = \sqrt{1+\frac{c^4 r_p^2}{G^2 M^2}}= \sqrt{1+4\frac{r_p^2}{r_s^2}} \simeq 2 \frac{r_p}{r_s} \gg 1.
\label{eccN2}
 \enq
In either relativistic or Newtonian theory the trajectory's shape is a rather extreme hyperbola
(see Goldstein \cite{Goldst} p. 94). Now the starting and ending points of the trajectory
are known: The photon trajectory starts at the star with the cylindrical coordinates $(r,\theta)=(r_\infty,0)$ and ends on earth with the cylindrical coordinates $(r,\theta)=(r_{earth},\pi)$. Thus according to \ern{Goldst}:
\beq
 0 = \frac{1}{r_\infty} = \frac{k}{l^2}[1+e \cos (\theta')] \Rightarrow
 e \cos (\theta') = -1 \Rightarrow \cos (\theta')= - \frac{1}{e}
 \label{thetp}
 \enq
 \beq
 \frac{1}{r_{earth}} = \frac{k}{l^2}[1+e \cos (\pi-\theta')] \Rightarrow
   \frac{1}{r_{earth}}= \frac{k}{l^2}[1-e \cos (\theta')] = \frac{2 k}{l^2}.
 \label{rear}
 \enq
 It follows that we may also estimate $l$ by:
 \beq
 l = \sqrt{2 k r_{earth}}.
 \label{lrear}
 \enq
We may insert the above results into \ern{Goldst} and write
\beq
 \frac{1}{r} = \frac{1}{2 r_{earth}}[1+e \cos (\theta - \theta')] \Rightarrow
 r = \frac{2 r_{earth}}{1+e \cos (\theta - \theta')}.
 \label{Goldste}
 \enq
We are now at a position to calculate the lensing angle given by \ern{lensangle}:
\beq
\tan \alpha = \frac{v_\bot}{v_\|} = \frac{dy}{dx} =
\frac{\frac{dy}{d \theta}}{\frac{dx}{d \theta}}
\label{lensangle2}
\enq
Now:
\beq
x (\theta) = r (\theta) \cos \theta, \quad y (\theta) = r (\theta) \sin \theta.
\label{xy}
\enq
Calculating the derivative of the above quantities we have:
\beq
\frac{dx}{d \theta}  = \frac{dr}{d \theta} \cos \theta - r \sin \theta, \qquad
\frac{dy}{d \theta}  = \frac{dr}{d \theta} \sin \theta + r \cos \theta.
\label{dxy}
\enq
The quantity $\frac{dr}{d \theta}$ is deduced from \ern{Goldste}
\beq
  \frac{dr}{d \theta}  = \frac{2 r_{earth} e \sin (\theta - \theta')}{[1+e \cos (\theta - \theta')]^2}.
 \label{drdth}
 \enq
Inserting \ern{drdth} into \ern{dxy} and using some basic trigonometric identities and
\ern{thetp} leads to the following results:
 \beq
\frac{dx}{d \theta}  = \frac{-2 r_{earth} }{[1+e \cos (\theta - \theta')]^2}
[e \sin \theta'+ \sin \theta]
, \qquad
\frac{dy}{d \theta}  = \frac{2 r_{earth} }{[1+e \cos (\theta - \theta')]^2}
[ \cos \theta - 1].
\label{dxy2}
\enq
It is now easy to insert the expression of \ern{dxy2} into \ern{lensangle2} and obtain
a simple expression:
\beq
\tan \alpha = \frac{1-  \cos \theta}{e \sin \theta'+ \sin \theta}.
\label{lensangle3}
\enq
Thus, at the star $\theta = 0$ and thus the light ray satisfies $\tan \alpha = 0$, with a plausible
physical solution of $\alpha = \pi$. This means that initially the light ray is propagating parallel to the x-axis in the direction of earth. On the other hand as the light ray approaches earth, we have $\theta = \pi$ leading to:
\beq
\tan \alpha = \frac{2}{e \sin \theta'} = \pm \frac{2}{e \sqrt{1-\frac{1}{e^2}}}
\simeq \pm \frac{2}{e},
\label{lensangle4}
\enq
In which we used the fact that $e \gg 1$ as follows from \ern{eccE2} and \ern{eccN2}. Choosing
the physical plausible positive sign and taking into account that the lensing angle is small we have:
\beq
 \alpha \simeq \frac{2}{e} \Rightarrow \alpha_E \simeq \frac{2}{e_E} = \frac{2 r_s}{r_p} = 2 \alpha_N.
\label{lensangle4b}
\enq
The Schwarzschild radius of the sun can be calculated from \ern{Schwaradius} from the sun's mass which is $M_{sun} \simeq 1.99 \ 10^{30} \ {\rm kg} $ leading to $r_s \simeq 2950 \ {\rm m}$. $r_p$ was taken in the classical observation of Eddington \cite{Edd} to be the Sun's radius, thus $r_p = 6.96 \ 10^8 \ {\rm m} $, it now follows that:
\beq
\alpha_E \simeq 8.47 \ 10^{-6} \ {\rm radians} = 1.75 \ {\rm arcseconds},
\label{lensangle5}
\enq
\beq
\alpha_N \simeq 4.27 \ 10^{-6} \ {\rm radians} = 0.87 \ {\rm arcseconds}.
\label{lensanglN5}
\enq
Eddington was able to show during a sun eclipse that the observed lensing angle is
closer to $\alpha_E$ than to $\alpha_N$ thus adding an important empirical argument
if favour of general relativity in addition to the theoretical arguments presented by
Einstein.

To conclude this section we notice that the maximal $h_{00}$ in the solar system is
according to \ern{hSchwa}:
\beq
|h_{00 \ max}|=  \frac{r_s}{r_p} = \frac{1}{2} \alpha_E = \frac{1}{e_E} \simeq 4.27 \ 10^{-6} .
\label{hSchwammax}
\enq
Thus justifying the linear approximation of \ern{lg} easily at least for the Solar system.

\section{A suggested experiment}

\begin{figure}[H]
\centering
\includegraphics[width=\columnwidth]{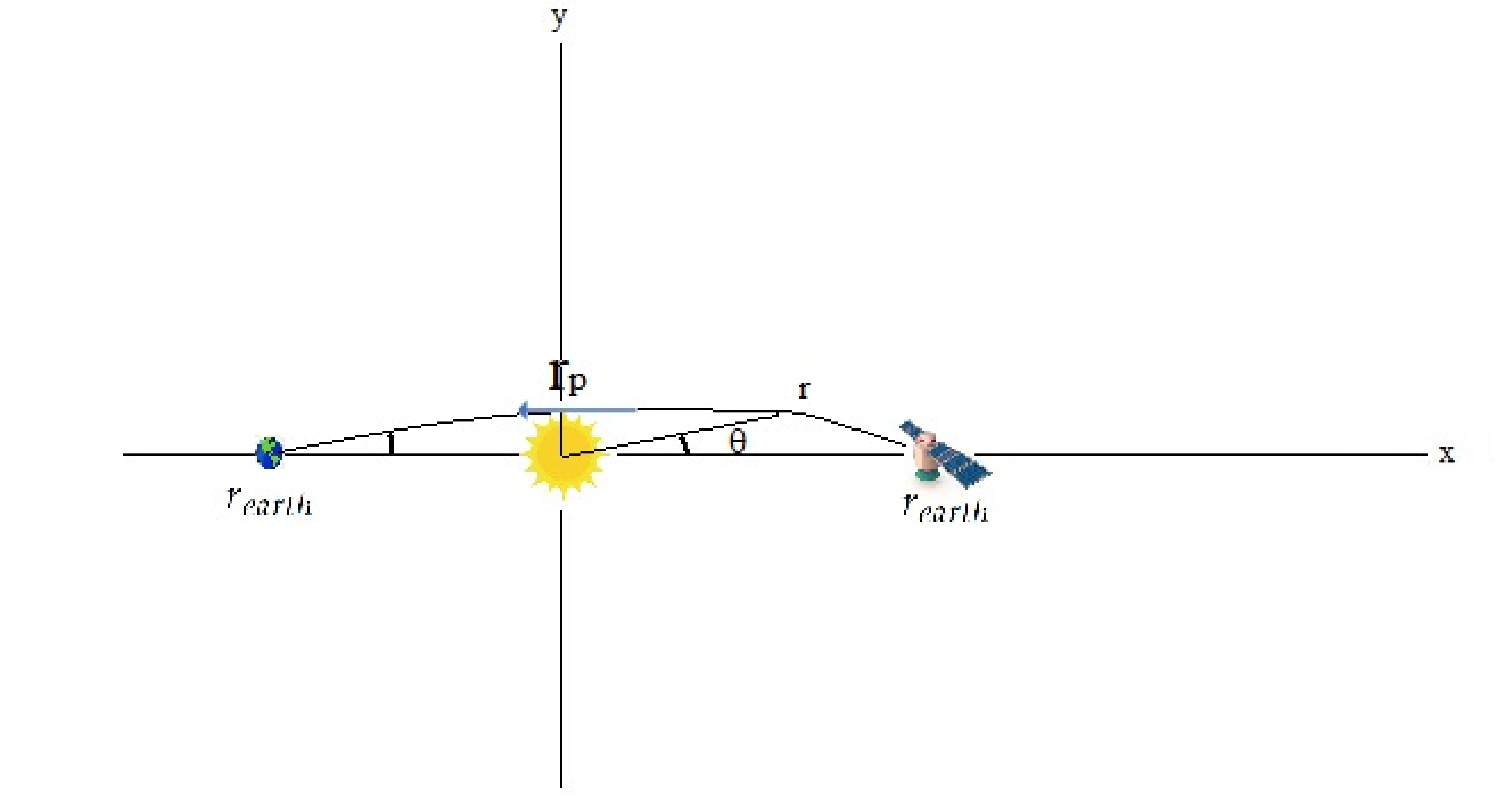}
 \caption{The trajectory of light from an antipodal satellite to the earth as affected by the sun's gravity}
 \label{trasat}
\end{figure}

The lensing angle measured by Eddington is not the only possible lensing angle in the Solar system.
To see that other angles are possible consider the case depicted in figure \ref{trasat}.
Here, the light ray is emitted from a satellite on the same trajectory as the earth but in the other side of the sun. Consider a line connecting the earth and the satellite, and assume that this line coincides with the x-axis. The sun is located at the origin of axis and both the satellite and the earth are in equal distance from the sun.
Now the starting and ending points of the trajectory
are known also in this case: The photon trajectory starts at the satellite with the cylindrical coordinates $(r,\theta)=(r_{earth},0)$ and ends on earth with the cylindrical coordinates $(r,\theta)=(r_{earth},\pi)$. Thus according to \ern{Goldst}:
\beq
 \frac{1}{r_{earth}} = \frac{k}{l^2}[1+e \cos (\theta')]
 \label{thetps}
 \enq
 \beq
 \frac{1}{r_{earth}} = \frac{k}{l^2}[1+e \cos (\pi-\theta')] \Rightarrow
 \frac{1}{r_{earth}} = \frac{k}{l^2}[1-e \cos (\theta')].
 \label{rear2sat}
 \enq
The difference of the above two equations leads to:
\beq
  2 e \cos (\theta') = 0 \Rightarrow \theta' = \frac{\pi}{2}.
 \label{thetps2}
 \enq
And their sum leads to:
\beq
 \frac{2}{r_{earth}} = \frac{2 k}{l^2} \Rightarrow \frac{1}{r_{earth}} = \frac{k}{l^2}.
 \label{rear2satb}
 \enq
Hence $l$ can be also evaluated as:
\beq
 l = \sqrt{k r_{earth}}.
 \label{lrear23}
 \enq
We may insert the above results into \ern{Goldst} and write:
\beq
 \frac{1}{r} = \frac{1}{r_{earth}}[1+e \sin \theta] \Rightarrow
 r = \frac{r_{earth}}{1+e \sin \theta}.
 \label{Goldste23}
 \enq
The quantity $\frac{dr}{d \theta}$ is deduced from \ern{Goldste23}
\beq
  \frac{dr}{d \theta}  = -\frac{ r_{earth} e \cos \theta }{[1+e \sin \theta]^2}
  =  -\frac{ r e \cos \theta }{1+e \sin \theta}.
 \label{Goldste232}
 \enq
Inserting \ern{Goldste23} and \ern{Goldste232} into \ern{dxy} will lead after some trigonometry to:
\beq
\frac{dx}{d \theta}  = \frac{-r}{1+e \sin \theta}[e + \sin \theta]
, \qquad
\frac{dy}{d \theta}  = \frac{r  \cos \theta}{1+e \sin \theta}.
\label{dxy2sat}
\enq
It is now easy to insert the expression of \ern{dxy2sat} into \ern{lensangle2} and obtain
a simple expression:
\beq
\tan \alpha = \frac{-  \cos \theta}{e + \sin \theta}.
\label{lensangle3b}
\enq
The ray should be launched from $(r,\theta)=(r_{earth},0)$ at an angle:
\beq
\tan \alpha = \frac{-1}{e}.
\label{lensangle4c}
\enq
that is at according to \ern{eccE2} a small angle to the negative x direction of $\frac{1}{e}$. And will arrive at earth at the angle:
\beq
\tan \alpha = \frac{1}{e}.
\label{lensangle4d}
\enq
that is at a small angle $\frac{1}{e}$ below the negative x axis, causing the observer to see a deviation of the location of the satellite of the same angular magnitude, which is:
\beq
\alpha_E \simeq \frac{1}{e_E} = 4.27 \ 10^{-6} \ {\rm radians} = 0.87 \ {\rm arcseconds},
\label{lensangle5sat}
\enq
\beq
\alpha_N \simeq \frac{1}{e_N} = 2.13 \ 10^{-6} \ {\rm radians} = 0.43 \ {\rm arcseconds}.
\label{lensanglN5sat}
\enq
The relativistic $\alpha_E$ is expected rather than the Newtonian $\alpha_N$. Notice that in this
case the angular deviation is half of the one expected for a distant star.
Of course one can use a laser source in a short wavelength that is not common in the suns
natural radiation and modulate the laser such that a matched filter can be easily constructed
in the receiver side to minimize noise interference.

\section {Beyond the Newtonian Approximation}

So far we have considered only the case of Newtonian potential which neglects retardation phenomena.
This approach seem to suffice for the solar system in which despite the sun's slow change in mass through the solar wind, retardation effect seem to be negligible.

However, in other gravitating systems such as galaxies (not to mention galaxy clusters) there mere size and the nature of their mass exchange with the environment leads to a situation in which retardation effects cannot be neglected. Indeed it was shown \cite{YaRe3} that the peculiar shape of galactic rotation curves can be explained by the retardation phenomena. This raises the question on the effect of the same on gravitation lensing. Let us reiterate our main results. The velocity of a photon in a gravitational field is given to the first order of $h_{00}$ according to \ern{photonequapar4}, \ern{photonequa2d}, and \ern{phi}:
\beq
  v_\| = c ( h_{00} + 1), \quad \frac{d \vec v_\bot }{d t}  =
  -c^2 \vec \nabla_\bot h_{00} =  -2 \vec \nabla_\bot \phi, \quad
  \phi = \frac{c^2}{2} h_{00} = -G \int \frac{\rho (\vec x', t-\frac{R}{c})}{R} d^3 x'
 \label{photonequastartpoint}
\enq
The duration $\frac{R}{c}$ for galaxies may be a few tens of thousands of years, but~can be considered short in comparison to the time taken for the galactic density to change significantly. Thus, we can write a Taylor series for the density:
\beq
\rho (\vec x', t-\frac{R}{c})=\sum_{n=0}^{\infty} \frac{1}{n!} \rho^{(n)} (\vec x', t) (-\frac{R}{c})^n,
\qquad \rho^{(n)}\equiv \frac{\partial^n \rho}{\partial t^n}.
\label{rhotay}
\enq

By inserting Equations~(\ref{rhotay}) into Equation~(\ref{photonequastartpoint}) and keeping the first three terms, we will obtain:
\beq
\phi = -G \int \frac{\rho (\vec x', t)}{R} d^3 x' +  \frac{G}{c}\int \rho^{(1)} (\vec x', t) d^3 x'
- \frac{G}{2 c^2}\int R \rho^{(2)} (\vec x', t) d^3 x'
\label{phir}
\enq

The Newtonian potential is the first term, the~second term contributes only to $v_\|$ but no to
$\vec v_\bot$ , and~the third term is the lower order correction to the Newtonian potential affecting $\vec v_\bot$ in \ern{photonequastartpoint} :
\beq
\phi_r = - \frac{G}{2 c^2} \int  R \rho^{(2)} (\vec x', t) d^3 x'
\label{phir2}
\enq

We recall that according to \ern{lensangle} it is only $\vec v_\bot$ that affects the lensing angle,
and thus despite the fact that second term $\frac{G}{c}\int \rho^{(1)} (\vec x', t) d^3 x'$ has a physical measurable effect on $ \vec v_\|$ it does not have any effect on the lensing phenomena
to the first order in $h_{00}$.

We underline again that we are {\bf not} considering a post-Newtonian approximation in this paper
(see section \ref{linear}), in~which matter travels at nearly relativistic speeds, but~
we will be considering the retardation effects and finite propagation speed of the gravitational field. We emphasize that taking $\frac{v}{c} \simeq 10^{-3} << 1$ (as for galaxies $v \simeq 10^5 \ {\rm m/s}$ see figure \ref{vcrhoc2} while $c \simeq 3 \ 10^8 \ {\rm m/s}$)
 is not the same as taking $\frac{R}{c} << t_r$ (with $R$ being the typical size of a galaxy say about: $R \simeq 3 \ 10^{20} \ {\rm m} \Rightarrow \frac{R}{c} \simeq \ 10^{12} \ {\rm s}$) and $t_r$ is the typical time the mass of the same galaxy changes (to be discussed in section \ref{dynmodel}).
\Ern{photonequa2d} can be rewritten in terms of a perpendicular "force" per unit mass such that:
\beq
 \frac{d \vec v_\bot }{d t}  = \vec F_\bot, \qquad \vec F_\bot \equiv  -2 \vec \nabla_\bot \phi
 \label{photonequa2df}
\enq
The total perpendicular force per unit mass is:
\ber
\vec F_\bot &=& \vec F_{N \bot} + \vec F_{r \bot}
\nonumber \\
 \vec F_{N \bot} &=& - 2 \vec \nabla_\bot \phi_N =  - 2 G  \int \frac{\rho (\vec x',t)}{R^2} \hat R_\bot d^3 x',
 \qquad \hat R \equiv \frac{\vec R}{R}, \quad
 \hat R_\bot \equiv \hat R -  \hat v_0 (\hat v_0 \cdot \hat R)
\nonumber \\
 \vec F_{r \bot} &\equiv& - 2\vec \nabla_\bot \phi_r =  \frac{G}{c^2} \int  \rho^{(2)} (\vec x', t) \hat R_\bot d^3 x'
\label{Fr}
\enr
Now if the lensing trajectory is far from the gravitating mass such that
$r \equiv |\vec x| \gg r' \equiv |\vec x'|$ it follows that $R \simeq r$ and
$\hat R \simeq \hat r \equiv \frac{\vec x}{|\vec x|}$. The force can now be written
as:
\ber
\vec F_\bot &=& \vec F_{N \bot} + \vec F_{r \bot}
\nonumber \\
 \vec F_{N \bot} &\simeq&  - 2 G  \int \frac{\rho (\vec x',t)}{r^2} \hat r_\bot d^3 x'
  = - \frac {2 G M}{r^2} \hat r_\bot ,
 \qquad  \hat r_\bot \equiv \hat r -  \hat v_0 (\hat v_0 \cdot \hat r)
\nonumber \\
 \vec F_{r \bot} &\simeq&   \frac{G}{c^2} \int  \rho^{(2)} (\vec x', t) \hat r_\bot d^3 x'
= \frac{G}{c^2} \ddot{M} \hat r_\bot.
\label{Fr2}
\enr
In the above $M = \int \rho d^3 x'$ is the total mass of the gravitating body, which may be a galaxy or a cluster of galaxies, and $\ddot{M}$ is the second derivative of the same. Now
although $\ddot{M}$ may be affected by many astrophysical processes we suggest that main contribution is the depletion of gas outsider the gravitating body (see section for
a detailed discussion in section \ref{dynmodel}), hence $\ddot{M} = - |\ddot{M}| < 0$.
It follows that according to \ern{Fr2}:
\beq
\vec F_\bot \simeq - [\frac {2 G M}{r^2} -\frac{G}{c^2} \ddot{M}]  \hat r_\bot
=  - \frac {2 G (M+M_d (r))}{r^2} \hat r_\bot
\label{Fr3}
\enq
where the "dark matter" mass is defined as:
\beq
M_d (r) \equiv  \frac{r^2 |\ddot{M}|}{2  c^2}
\label{Fr40}
\enq
we observe that the "dark matter" mass associated with lensing is the same as
the "dark matter" mass associated with galactic rotation curves (see equation (106) of \cite{YaRe3}), thus explaining the observational results of \cite{[54]}.

\section{A Dynamical~Model}
\label{dynmodel}

As~mass is accumulated in the galaxy or galaxy cluster, it must be depleted in the surrounding medium.
This is due to the fact that the total mass is conserved; still, it is of interest to see if this intuition is compatible with a model of gas dynamics. For~simplicity, we assume that the gas is a barotropic ideal fluid and its dynamics are described by the Euler and continuity equations as follows:
\beq
\frac{\partial{\rho}}{\partial t} + \vec \nabla \cdot (\rho \vec v ) = 0
\label{masscon}
\enq
\beq
\frac{d \vec v}{d t} \equiv
\frac{\partial \vec v}{\partial t}+(\vec v \cdot \vec \nabla)\vec v  = -\frac{\vec \nabla p (\rho)}{\rho} - \vec \nabla \phi
\label{Euler}
\enq
where the pressure $p (\rho)$ is assumed to be a given function of the density,  $\frac{\partial }{\partial t}$
is a partial temporal derivative, $\vec \nabla$ has its standard meaning in vector analysis and $\frac{d }{d t}$ is the material temporal derivative. We have neglected viscosity terms due to the low gas density.

\subsection{General considerations}

Let us now take a partial temporal derivative of \ern{masscon} leading to:
\beq
\frac{\partial^2 {\rho}}{\partial t^2} + \vec \nabla \cdot (\frac{\partial{\rho}}{\partial t} \vec v +
 \rho \frac{\partial \vec v}{\partial t} ) = 0.
\label{masscon2b}
\enq
Using \ern{masscon} again we  obtain the expression:
\beq
\frac{\partial^2 {\rho}}{\partial t^2} = \vec \nabla \cdot \left(\vec \nabla \cdot (\rho \vec v ) \vec v  -
 \rho \frac{\partial \vec v}{\partial t} \right) .
\label{masscon3}
\enq
We divide the left and right hand sides of the equation by $c^2$ as in \ern{Fr} and obtain:
\beq
\frac{1}{c^2}\frac{\partial^2 {\rho}}{\partial t^2} = \vec \nabla \cdot \left( \frac{\vec v}{c} \left(\frac{\vec v}{c} \cdot \vec \nabla \rho +
\rho \vec \nabla \cdot (\frac{\vec v}{c} )\right)   -
 \rho \frac{1}{c} \frac{\partial \frac{\vec v}{c}}{\partial t} \right) .
\label{masscon4}
\enq
Since $\frac{\vec v}{c} $ is rather small in galaxies and galaxy clusters it follows that $\frac{1}{c^2}\frac{\partial^2 {\rho}}{\partial t^2}$ is also small unless the density or the velocity have significant spatial derivatives. A significant acceleration $\frac{\partial \frac{\vec v}{c}}{\partial t}$ resulting from a considerable force can also have a decisive effect. The depletion of available gas can indeed cause such gradients as we describe below using a detailed model. Taking the volume integral of the left and right hand sides of \ern{masscon4} and using Gauss theorem we arrive at the following equation:
\beq
\frac{1}{c^2}\ddot{M}=\frac{1}{c^2} \int \frac{\partial^2 {\rho}}{\partial t^2} d^3 x = \oint d \vec S \cdot \left( \frac{\vec v}{c} \left(\frac{\vec v}{c} \cdot \vec \nabla \rho + \rho \vec \nabla \cdot (\frac{\vec v}{c} )\right)   -
 \rho \frac{1}{c} \frac{\partial \frac{\vec v}{c}}{\partial t} \right) .
\label{masscon5}
\enq
The surface integral is taken over a surface encapsulating the galaxy or galaxy cluster. This leads according to \ern{Fr40} to a "dark matter" effect of the form:
\beq
M_d (r)= \frac{r^2}{2} \oint d \vec S \cdot \left( \frac{\vec v}{c} \left(\frac{\vec v}{c} \cdot \vec \nabla \rho + \rho \vec \nabla \cdot (\frac{\vec v}{c} )\right)   -
 \rho \frac{1}{c} \frac{\partial \frac{\vec v}{c}}{\partial t} \right) .
\label{masscon6}
\enq
Thus we obtain the order of magnitude estimation:
\beq
\left[ \frac{M_d (r)}{M} \right] \approx \left( \frac{v}{c}\right)^2 \left[  \frac{r}{l_\rho}
 + \frac{r}{l_v}  + \frac{r}{l_d} \right]
\label{masscon7}
\enq
In the above we define three gradient lengths:
\beq
l_\rho \equiv \frac{\rho}{|\vec \nabla \rho|}, \qquad
l_v \equiv \frac{v}{|\vec \nabla \cdot v|}, \qquad
l_d \equiv \frac{v^2}{|\partial_t v|}
\label{masscon8}
\enq
We can also write:
\beq
\frac{1}{l_t}  =\left[  \frac{1}{l_\rho}  + \frac{1}{l_v}  + \frac{1}{l_d} \right]
\label{masscon9}
\enq
in which the smallest gradient length will be the most significant one in terms of
the "dark matter" phenomena. In the depletion model to be described below we assume
that $l_\rho$ associated with density gradients is the shortest length scale.
For galaxies we have $\left( \frac{v}{c}\right)^2 \approx 10^{-6}$, hence the factor
$\frac{r}{l_t}$ should be around $10^{6}$ to have a significant "dark matter" effect.
A detailed model of the depletion process in galaxies leading to the desired second derivative
of galactic mass is given in \cite{YaRe3} and will not be repeated here.

\section{Conclusions}

In this paper we have deduced from general relativity a linear approximation. Under the said linear
approximation we have solved Einstein field equations in term of retarded solutions. Those where
used to derive the trajectory of a light ray (photon) both in the direction parallel to its original direction and perpendicular to it. Our current approach was compared to the approach based on the  Schwarzschild metric \cite{Weinberg} and was shown to be better in the sense that it is applicable to general mass distributions including ones that are changing in time. The light ray equations is the presence of a static point mass allowed us to re-derive the classical light deflection  of Einstein and Eddington \cite{[12]} and suggest a new experiment. This was followed by a discussion on the retardation effects on light ray trajectories, deriving an expression for "dark matter" which is the same as the one obtained in
\cite{YaRe3} for slowly moving bodies. Thus justifying results reported in the literature of the equivalence of "dark matter" for both galaxy rotation curve and gravitational lensing. This is followed by a discussion on the physical requirements needed in order to derive "dark matter" effects from retardation and a detailed model showing how those requirements are satisfied in a galactic scenario.

Lorentz symmetry invariance does not allow action at a
distance potentials and forces, but retarded solutions are allowed. Retardation is significant for large distances and large second derivatives. It should be emphasized that the retardation approach does not require that velocities in, $v$ in the gravitating body are high; in fact, galactic \& galactic cluster bodies (stars, gas) move slowly with respect to the speed of light—thus the~quantity $\frac{v}{c} \ll 1$. Typical velocities in galaxies are $100~{\rm \frac{km}{s}}$ (see Figure~\ref{vcrhoc2}), which makes $\frac{v}{c}$ to be about $0.001$ or smaller.
However, every gravitational system, even if it consists of subluminal entities, has a retardation distance, above~which retardation cannot be neglected.  Natural systems, for example a star or a galaxy and even a galactic cluster, exchanges mass with its environment. The~sun loses mass through solar wind and galaxies accrete matter which originate in the intergalactic medium. Thus all natural (gravitational) systems have a finite retardation distance. This leads to quantitative inquiry: what is the actual size of the retardation distance? The modification of the solar mass is quite small and thus the retardation distance of the solar system is extremely large, we can thus neglect retardation within the solar system. On the other hand, for~the M33 galaxy, velocities indicate that retardation cannot be neglected. The retardation distance was calculated in \cite{YaRe3} to be roughly $R_r = 4.54$ kpc for M33; other galaxies of different types have shown similar results~\cite{Wagman}. We demonstrated,  in~Section~\ref{dynmodel},  that~this does not require a high velocity of gas or stars and is perfectly consistent with current observational knowledge of galactic and extragalactic dynamics.

We underline,  that if~extra galactic mass is abundant (or totally consumed), $\ddot{M} \simeq 0$ and the retardation force vanish. As was reported~\cite{Dokkum} for~NGC1052-DF2.

We emphasize that the terms in the GR equations responsible for gravitational radiation recently discovered are also the cause for the peculiar shape of the rotation curves of galaxies and gravitational lensing. The approximation used here is not a far field approximation but a near field one. Indeed, the~expansion ~(\ref{phir}), being second order, is only valid up to limited~radii:
\beq
R < c \ T_{max} \equiv R_{max}
\label{Rmax}
\enq

This is reasonable since the extension of the rotation curve in galaxies and the distance of lensing trajectories from galaxies
is the same order of magnitude as the size of the galaxy. The case in which the dimensions of the source is much smaller than the distance to the observer will result in a different valid approximation to (\ref{bhint}), leading to the famous quadruple expression of gravitational radiation, as~derived by Einstein~\cite{Einstein2} and verified (indirectly) in 1993 by Russell A. Hulse and Joseph H. Taylor. The~observation of the Hulse–Taylor binary pulsar has given the first (indirect) evidence of gravitational waves~\cite{Taylor}. On~11 February 2016, the~LIGO
and Virgo Collaboration announced that they made the first (direct) observation of gravitational waves. The~observation was made earlier, on~14 September 2015, using the LIGO detectors. The~gravitational waves were caused by the merging of a binary black hole system~\cite{Castelvecchi}. Thus, here we discuss only a near-field application of gravitational radiation contrary to previous works discussing far- field~results.

Unfortunately no direct measurement of  the second temporal derivative of the galactic mass is  available. What is available is the remarkable fit between the retardation theoretical velocity and the observed galactic rotation curve, as~can be seen in Figure~\ref{vcrhoc2}, this constitutes indirect evidence of the total mass second derivative. Competing theories like dark matter do not supply any direct observational evidence either. Despite the work of many people and the investment of a large financial resources, there is no evidence of dark matter. Occam's razor postulates that when theories compete, the~one that makes less ontological assumptions about the physical existence of an exotic form of matter, wins. Retardation theory assumes only baryonic matter and a large second temporal derivative of mass.

Problems related to dark matter, such as the core-cusp problem which refers to the difference between the dark matter density profiles of galaxies and the density profiles predicted by N-body simulations. Almost~all simulations form dark matter halos, which posses "cuspy" dark matter mass distributions, with~density increasing rather steeply at small radii, while the rotation curves of most dwarf galaxies suggest that they have flat central dark matter profiles ("cores"). This does not occur in the retardation theory which does not require dark matter. One cannot consider flat or sharp profiles of dark matter distribution if dark matter is not there. The~persistent difficulties with dark matter's dynamics strengthen the claim that dark matter is not needed and the gravitational lensing characteristics attributed to dark matter should be attributed to~retardation.

To conclude, we would like to mention the significant theory of conformal gravity put forward by Mannheim~\cite{Mannheim1,Mannheim2}.
The current retardation approach leads approximately to a Newtonian potential and in addition a linear potential. Such potential types can be derived from
 conformal gravity. On~purely phenomenological grounds, rotation curve fits for linear plus Newtonian  potentials have already been published. While those fits are very good, they had to treat the coefficient of the linear potential as a variable that changed from galaxy to galaxy; this element contradicts conformal gravity in
which the coefficient of the linear potential is a universal constant. This can be explained
in the framework of retardation, in~which  a $\ddot{M}/M$ depends on the dynamical particular conditions of every galaxy. Indeed, $\ddot{M}/M$ has a theoretical reason, lacking in a pure phenomenological approach. The work of Mannheim~\cite{Mannheim0,Mannheim1,Mannheim2}, is related to conformal gravity, which is different from GR, and~thus has to justify other results of GR (Big Bang Cosmology, etc.). We also underline that retardation per se does not contradict conformal gravity and, both effects may exist, although~ the principle of Occam's razor forbids us to add new universal constants if the existing ones suffice to explain~observations.

Retardation theory's approach is minimalistic (it satisfies the Occam's razor  rule), and does not affect observations that are beyond the near-field regime and, thus, does not contradict GR theory and its observational consequences (nor with Newtonian theory, as~the retardation effect is negligible for "small" distances). The perfect fit to the rotation curve and gravitational lensing is achieved with a single parameter and we do not adjust the mass to light ratio in order to improve our fit as is done by other authors.
Retardation effects beyond gravity, in particular with respect to electromagnetic theory were studied in~\cite{Tuval,YahalomT,Yahalom3,Yahalom4}.

In this paper we study the gravitational lensing scenario showing that retardation effects
lead to the same "dark matter" mass for lensing as for galactic rotation curves.

The current paper does not discuss dark matter in a cosmological context; this is left for future work. We mention, however, that on  the cosmological scales it is not enough to invoke dark matter, but one must also consider dark energy as described in the $\Lambda$CDM model.

The CMB anisotropy spectrum that was observed precisely using WMAP in 2003-2012, and even to higher precision by the Planck spacecraft in 2013-2015 is in agreement with the $\Lambda$CDM model \cite{Hinshaw,Ade}.

Moreover, small anisotropies of the homogenous universe grew gradually and condensed the homogeneous material into stars, galaxies and larger structures. Since ordinary matter is affected by radiation, which is dominant at very early times. It follows,that its density perturbations are washed out and unable to condense \cite{Jaffe}. Thus it is suggested that if there is only baryonic matter, there would not have been sufficient duration for perturbations to grow into  galaxies and clusters seen.

Dark matter is assumed to provides a solution to this problem because it does not interact with electromagnetic radiation. Therefore, its perturbations can grow fast. The resulting gravitational potential attracts ordinary matter collapsing later, speeding significantly the structure formation process \cite{Jaffe,Low}.

On the other hand we know \cite{YaRe3} that the Newtonian gravity is only a part of the gravitational force and at large distances retardation forces prevail. Thus what is attributed to
dark matter attraction in terms of density perturbation growth may be attributed to retardation.

The CAMB anisotropy spectrum as well as the distant Super Novae data may be
explained by a more thorough perturbation analysis of the Friedman Robertson Walker metric,
preliminary results in this direction are given in \cite{Yahalomd}.

\authorcontributions{This paper has a single author, which has done all the work presented.}

\funding{This research received no external~funding.}

\acknowledgments{The author wishes to thank his former student, Michal Wagman, for~supplying the data points for the rotation curve of the M33 galaxy.
This work is a result of more than twenty five years of thinking (discontinuously) on the dark matter problem, which was first suggested to
me by the late Jacob Bekenstein during my stay at the Hebrew University of Jerusalem. The~current retardation solution arose from my discussions with the late Donald Lynden-Bell of Cambridge University, and~the late Miron Tuval. Other people with which I discussed this work and offered important feedback are Lawrence Horwitz of Tel-Aviv University and  Marcelo Shiffer of Ariel University.
This work benefitted from discussions with  James Peebles, Neta Bachall and Sam Cohen, all from Princeton University.
Special thanks are due to   Jiri Bicak for an invitation to present the theory at Charles University in Prague. I would like to thank
Philip Mannheim and   James Obrien
for our discussions during the recent IARD meetings and for supplying some relevant data. I have benefited from discussions with a long list of distinguished scientists and ask their forgiveness for not mentioning them~all.}

\conflictsofinterest{The author declares no conflict of~interest.}

%%%%%%%%%%%%%%%%%%%%%%%%%%
\reftitle{References}

\publishersnote{MDPI stays neutral with regard to jurisdictional claims in published maps and institutional affiliations.}
\end{document}